# Learning Analytics: A Survey

Usha Keshavamurthy[#1], Dr. H S Guruprasad[*2]

*# PG Scholar, Dept. of CSE, BMSCE*
*\* Professor and Head, Dept. of CSE, BMSCE*
*BMSCE, Bangalore, India*

*Abstract*— Learning analytics is a research topic that is gaining increasing popularity in recent time. It analyzes the learning data available in order to make aware or improvise the process itself and/or the outcome such as student performance. In this survey paper, we look at the recent research work that has been conducted around learning analytics, framework and integrated models, and application of various models and data mining techniques to identify students at risk and to predict student performance.

*Keywords*— Learning Analytics, Student Performance, Student Retention, Academic analytics, Course success.

## I. INTRODUCTION

Learning Analytics is a research field related to Educational Data Mining (EDM), and is gaining increasing popularity since Horizon Report 2012 described it as a forthcoming trend. Learning analytics (LA) refers to the application of Big Data methodologies and techniques to improvise the learning. LA is based on analyzing the learning behavior of students using a wide data set that takes into consideration- the student enrollment data, the previous academic record of students, student surveys through questionnaires about courses and teaching methods, data from online discussion forums and such.

In predicting and analyzing student performance, there are a no. of techniques that could be used such as classification algorithms - Eg: decision tree methods- C4.5, RepTree and J48, k-nearest neighbor classifier, Naive Bayes, Multi-layer perceptron (neural networks), Sequential Minimal Optimization and clustering methods - Eg: Latent Semantic Analysis and K-means clustering methods.

There could be various reasons for a student to fail or drop out from the course such as financial issues, health problems, family issues, time-management, under-preparedness etc. The educational institutions are, however, interested only in the predictor variables that come under the institutional purview or control which is lack of preparation on student front. If the institution is able to identify early, the students who would need help in completing their course, it can put forth some preventive actions such as customizing the course material for the targeted students or provide additional coaching. To this front, learning analytics plays a key role in predicting student performance which would benefit both the students and educational institutions.

In this survey paper, we look at the recent research work surrounding the field of learning analytics. Detailed survey follows in the next section.

## II. LITERATURE SURVEY

Ferguson Rebecca [1] provides a review of learning analytics. The paper examines various factors like technological, educational and political factors that drive learning analytics such as Big data, Online learning, political and economic concerns. The authors have described how data driven analytics developed and how learning analytics emerged, and the relationships between LA and educational data mining and academic analytics. Whereas educational data mining focuses on how to extract useful data from a large learning dataset, learning analytics focuses on optimizing opportunities in online learning environment. Academic analytics, on the other hand, focuses on improving learning opportunities and educational results across national and international levels. The authors also identify a future set of challenges to be addressed by learning analytics like establishing clear set of ethical guidelines, coupling with recent and emerging learning technologies, understanding learner perspectives and working with wide range of learning datasets. Wolfgang Greller et. al. [2] propose a generic framework for Learning Analytics that considers six critical dimensions, namely, Objectives, Data, Instruments, Internal Limitations, External Constraints and Stakeholders. The paper also touches upon the ethical perspective of learning analytics to protect the learners. Erik Duval [3] discusses capturing of the attention data in learning environment in a number of ways such as posts, comments and messages. The authors showcase how such attention metadata could be stored and used. There are two approaches explored: Learning dashboards to provide visual overview of user activities individually and in relation to their peers, and, Learning recommenders that can be used to filter and recommend resources based on user behavior data collected. Jie Zhang [4] propose a framework for course management system. With this framework, the student's usage of eLearning system along with access pattern can be studied with respect to time. Based on the predicted results, the course manager may customize the material for individual students, determine effective learning methods and also come to know the preferred learning devices. The framework mainly comprises of Data Infrastructure and Data Analytical modules. Data Infrastructure module makes use of Hadoop framework for distributed computation, distributed data storage and Data Broker service. Data Analytical modules aid to collect data from cloud, to adapt, refine and optimize data analytics flow and to mine usage patterns. The authors also describe the use-cases for the proposed framework. Alyssa Friend Wise et. al. [6] investigate on how students contribute and reciprocate to messages in online discussions in learning environment. The analytics is both embedded in learning environment and extracted from it, leading to an integrated model. The implementation uses an online wiki-based reflective journal shared between each student and course instructor. The reflective journal is used to record the student goals and periodic analytic information (such as class average score) along with student responses to a series of reflective questions. The initial findings showed that the reflective journal contributed to a productive class environment and participation by students. A valuable outcome of the findings was the invisible activity validation. Eg: ability to capture listening data such as people who were engaged intensely in discussions but did not post many comments and also the voracious speakers who had a need to improve on their listening efforts. Tim Rogers et. al. [9] have done a comparative study on index method v/s linear multiple regression method on Student data to identify students at risk of failure. The study showed that the correlation coefficients for both regression ($r = 0.70$) and index method ($r = - 0.58$) were significant and largely in line with each other. Another observation was that, the Index method, though with relatively lower correlation, had lesser variability in predictions which is an important consideration in prediction analysis. Sharon Slade et. al. [30] discusses how learning analytics enable Educational Institutions to gain knowledge on their Students Learning Behavior and use it to improve the rate of Student Retention and perform in-time interventions to drive Student Success. However, collection of student data required as input to predictive models face a number of ethical issues and challenges that has been the focus of this paper. Eg: Issues related to location and interpretation of data, privacy and transparency issues, impact of surveillance and management of data thus collected. This paper proposes a framework with a set of guiding principles to Educational Institutions for addressing such ethical issues in the field of Learning Analytics such as taking Informed Consent from students, and providing a choice to opt out from data collection.

Mohammed M. Abu Tair et. al. [13] discusses a case study conducted on fifteen year period graduate student dataset, to mine for knowledge from educational data. Data mining techniques have been applied to perform classification and clustering, identify associations and perform outlier analysis. Rule Induction and Naïve Bayesian classifier have been used for classification and K-means as clustering technique. The outlier analysis showed that the outliers were result of rare events and not due to errors. P K Srimani et. al. [14] have applied a number of data mining algorithms, namely, Bayesian classifier, decision table, MLP, J48, Ripper (rule based learner) on student data from elementary classes - I to VII to do performance analysis. In class I, Ripper was found most efficient and accurate. In class II, MLP was found as most efficient and accurate with 99.9% correctly classified instances. In class III and VI, all the algorithms performed equally well. For class IV, both J48 and Ripper performed efficiently. For class V also, most algorithms performed well. In case of class VII, MLP performed efficiently with accuracy of 99.8%. On considering the performance of various algorithms across all the classes, Bayesian algorithm outperformed the rest. Kabakchieva D [18] focus on applying Data Mining techniques to student data that includes student's personal, pre-university and university specific characteristics. Various classification algorithms such as Rule learners, decision tree technique, Bayes classifier and Nearest neighbor techniques have been used for analysis. The research work uses CRISP-DM (Cross-Industry Standard Process for Data Mining) model and WEKA software tool. The results showed that, the decision tree classifier performs best followed by rule learner and k-Nearest neighbor classifier method. Also, it is observed that Bayes classifiers are less accurate than the other methods used. However, it is noted that the prediction rates are not remarkable and varies between 52-67%. Edin Osmanbegovic et. al. [19] apply and compare three data mining techniques. viz., Bayesian classifier, neural networks and the decision tree method - J48. The research work makes use of WEKA software package and data collected from student surveys, students past success and present success. The impact of input variables has been analyzed using Chi-square, One-R, Info gain and Gain ratio tests. The results showed that GPA attribute impacted the output most followed by the attributes - entrance exam, study material and average weekly hours devoted to study. It was observed that Naive Bayes predicted better than the others and Multilayer Perceptron algorithm (neural networks) showed lowest prediction accuracy. V Ramesh et. al. [20] try to identify factors that influence student's performance and a suitable algorithm for predicting student grades. The dataset includes questionnaire data, and performance details collected. The study uses WEKA software implementation of the classifier algorithms - Naive Bayes, Multi-Layer Perceptron, Sequential Minimal Optimization, J48 (decision tree) and REPTree (decision tree) algorithms. Chi-square test was used to identify the significance between input variables used. It was noted that type of school (co-ed/boys/girls) did not influence grade, however, parent's education did have an influence on student grades. It was observed that MLP performed better with prediction accuracy more than 70%, in contrast to the findings made in [19]. Shaymaa E Sorour et. al. [21] predict the student performance using C (Current) method from the PCN method (Previous, Current and Next activity) and analyzed using Latent Semantic Analysis (LSA) and K-means clustering techniques. The C method studies only the student's free style comments data that are focused on student's class achievements and subject understanding. Mecab program is used to extract words and parts of speech after which LSA is applied to extracted words to identify patterns and relationships between the words and latent concepts. The results are then classified using K-means clustering technique. The results showed that for certain lessons, the student comments were good perhaps attributing to high motivation to express their attitude and write comments. The study expressed a correlation between self-evaluation comments by students and their academic performance.

Mohammed M. Abu Tair et. al. [10] apply Iterative Dichotomiser 3 (ID3) classification algorithm which uses decision tree concept to the student dataset for predicting pupils status. Dataset includes information like personal details, score details, extra-curricular activities. The study makes use of WEKA software tool to apply ID3 on the dataset. The result was prediction of which students in the dataset are eligible to apply for their Post Graduate courses. Brijesh Kumar Baradwaj et. al. [11] apply the decision tree technique ID3 to the dataset containing three years of information about student data. Based on Information Gain to decide the best attribute, PSM [Previous Semester Marks] attribute is selected as root of decision tree. Once all the data is classified, a number of IF-THEN rules are framed and validated against the formed decision tree to gain knowledge on student performance and provide resources to students who need special attention. Abeer Badr El Din Ahmed et. al. [12] apply the classification method of ID3 to five years of student dataset from an educational institution. Information Gain has been used to determine the best attribute. Mid-term attribute is found to have highest gain and hence placed at the root of decision tree. And once all the data is classified from the dataset, a number of IF-THEN rules are derived from the decision tree. The knowledge gained can be used to reduce student failure rate and do timely interventions. Adhatrao K et. al. [15] predict student performance by applying decision tree techniques - Iterative Dichotomiser 3 (ID3) and C4.5 classification algorithms using RapidMiner (data mining tool). Since the class labels - "Pass" / "Fail" are already known, the authors have decided to consider only the classification algorithms and not clustering techniques. The dataset uses student enrollment data, merit score in entrance exams and prior marks information (in XII) and admission type. As part of the study, the authors have developed a web application that can be used for prediction of individual performance as well as bulk performance. The results showed that both algorithms perform equally well and have comparable accuracy of above 75%. Ramanathan L et. al. [16] discusses that ID3 is found to be inclined towards attributes that have many values. Hence, the authors use modified ID3, that uses gain ratio instead of information gain and uses attribute weights at each step of decision making, implemented using C programming language. Other algorithms- J48 and Naïve Bayes are also applied using WEKA software and results are compared. The results showed that modified weighted ID3 (wID3) algorithm performed more efficiently compared to the other two with an accuracy of 93% while the other two had accuracy of around 75%.

Rebecca Barber et. al. [5] discusses a predictive analytical model created for the University of Phoenix, to identify students who are at the risk of failure. Two versions of logistic regression model have been developed. The initial model made use of demographic information, academic history, courses, credits, etc. Analyzing Model1 using SPSS randomization algorithm, prediction as to whether the student will pass or fail, was accurate more than 90% of the time. Model 2 used partial feed from an enterprise data warehouse environment. Using Naïve Bayes algorithm for analysis, Model 2 accurately predicted up to 95% of the time. Alfred Essa et. al. [7] propose an adaptive framework and a modeling strategy to address these limitations of generalization and interoperability. In a Student success system (S3), the advisor can login and view the list of his/her students along with risk indicators and further drill down to view student profiles or course-level activity and risks. S3 incorporates a number of visualizations to provide insights that are useful for diagnostic purposes. The idea behind the proposed modeling strategy is to enable selection of entire collection of hypotheses and be able to combine the predictions from multiple models, appropriately, and perhaps feed them as input to a higher prediction model. Cristobal Romero et. al. [8] propose a robust model for effective prediction of student's final performance based on data collected from online discussion forums and student profile data including prior performance history. The collected data goes through instance selection and attribute selection, after which different prediction techniques are applied, which includes a set of classification and classification via clustering techniques. The prediction models thus obtained from various techniques are then compared along with association rules mined. The results showed that it would help to do two predictions - one at the start and the other at the end of course and, using clustering techniques along with class association rules provide more interpretable models than using only the conventional classification methods. Also, using a set of relevant attributes instead of entire set of attributes and filtering only the forum messages relevant to the course content, helps in improving the classification accuracy. Mrinal Pandey et. al. [17] have applied various decision tree algorithms - J48, Naïve Bayes, Reptree and Simple cart using WEKA mining tool to predict Student Performance. The algorithm efficiencies are evaluated using cross validation and percentage split method. The results showed that J48 algorithm as the best among the ones compared, for model construction closely followed by Simple cart technique. Wolff Annika et. al. [23] have studied on how to predict students at risk in a distance learning setup as there is no face to face interaction between students and faculty. Here, the student behavior is observed by noting their access patterns in virtual learning environment (VLE) and compared against their previous behavior or the behavior of other students who have similar learning behavior, in order to predict students at the risk of failure. Cross validation method has been used to evaluate algorithm efficiency. The results showed that decision tree method - C4.5 outperformed the State Vector Machines (SVM) technique. Smith V C et. al. [24] discusses a case study for a community college offering online courses that makes use of learning analytics and predictive models to identify students at risk based on a number of variables. Data comprising of enrollment data, online student activity and grades is collected from RioLearn, a proprietary LMS and PeopleSoft system. Weighted student activities and standardized course lengths have been used for the study. Naive Bayes classifier has been used to make predictions and categorized into risk levels - low, moderate and high. The results showed a strong correlation between student activity and their outcome. Also, variables such as login frequency, site engagement, pace, assignment grades were effective for predicting course outcome. The authors have also demonstrated how the predictions can be coupled with other activity metrics to perform faculty interventions with positive results.

Jacob Kogan [22] discusses the Student Course Evaluation Questionnaire used as a conventional tool to determine Instructors teaching effectiveness, though, recent literature has found that certain factors may have an impact on Questionnaire responses. This paper analyzes the influence of variables such as class size, level, discipline and gender bias on student course evaluation. The results showed that student evaluations are better for faculty teaching larger classes compared to smaller classes, and student evaluations for female faculty was found to be better than those for male faculty and discipline (maths, science, etc.) plays an important role in student evaluations for faculty. Sotiris Kotsiantis et. al. [25] discusses a case study based upon blended learning approach using Moodle, an LCMS tool. Student perception and attitude towards learning tool and their interaction have been studied using different statistical and classification methods such as visualization, decision tree technique, class association rules and clustering methods. The results showed an association between student's negative attitude towards Moodle and their failure. On the other hand, there was also an association found between excellent academic grades and extensive use of LCMS tool. Kimberly E. Arnold et. al. [26] discusses Course signals (CS), which is a learning analytics tool used in Purdue University by the faculty members for timely interventions to improve Student success. The tool uses student enrollment data, past academic history, demographic characteristics and interaction with Blackboard Vista, LMS tool used by Purdue University. Real time feedback is provided to individual students by means of email by faculty that comes with a warning level indicating how the student is doing. The study shows that students using CS have higher retention and success rates compared to their better-prepared peers not using CS. The paper also provides student and instructors perception about the Course Signals tool. Both students and instructors believed that the tool is helpful and aids in the Institution's and Student's academic success. Doug Clow [27] discuss the issues and challenges faced in MOOC (Massive Online Open Courses) environment. Retention is a bigger problem with MOOCs compared to traditional education. Another challenge, is the lack of resources to circulate feedback and lack of support to students at risk. The authors coin the metaphorical phrase "funnel of participation" to describe the characteristics of MOOCs which shows a steep drop-off in the activity with course progress and unequal participation patterns. Three learning sites - iSpot, Cloudworks and OpenEd have been used to demonstrate the concept of funnel of participation. The funnel of participation calls out four major stages - Awareness (that MOOCs exist), Registration, Activity and Progress with large drop in numbers across the stages. However, there are online and distance teaching institutions with drop rates between conventional and MOOC systems. This shows that it is possible to mitigate the impact of the funnel. Ourania Petropoulou et. al. [28] present a new cloud-based analytics tool called LAe-R (Learning Analytics Enriched Rubric) which has been integrated with Moodle (LMS tool). The LAe-R tool is a blend that uses marking criteria and grades from conventional system as well as performance indicators gathered from learning behavior and access patterns in order to holistically assess student performance in e-learning environment. On review by practitioners, the tool received high ratings in regard to usability aspects and help content available with the tool. Giesbers B et. al. [29] study the relationship between student motivation and grades and participation and tool use in an online course that uses synchronous communication such as web-videoconferencing tool. This study affirms the relationship between student motivation and participation; and that between student motivation and grades. It shows a strong correlation between tool use and participation and also the association between the two and final scores. However, it is found that participation is a stronger predictor for final exam score than tool use.

### III. CONCLUSIONS

The survey covers several studies revolving around learning analytics - some proposing general framework to be used for learning analytics, one that presents a course management system and its use-cases, and several of them that apply and implement different mining algorithms to establish associations and predictions of student performance.

Quality data that translates to "efficiency of the data preprocessing (cleansing and transforming) steps" and relevance of data attributes are crucial in the model building and analysis phase to improve prediction accuracy. Visualization of student learning behavior through learning dashboards with respect to peers and bringing about awareness of student activities can aid in learning process. By studying the learning behavior, course facilitators can identify students at risk of failure and can take several initiatives such as customizing learning materials, providing additional coaching, recommending certain learning resources to help individual students at risk early in the learning process. Also, the survey shows that different experiments have yielded in different prediction accuracies and have contradictory comparison results in terms of algorithms used for predictive analysis. There is a need to establish common ground and gather sufficiently sized datasets to make concrete conclusions regarding the best suitable algorithm for making predictions related to student performance.

### ACKNOWLEDGMENT

The authors would like to acknowledge and thank Technical Education Quality Improvement Program [TEQIP] Phase 2, BMS College of Engineering and SPFU [State Project Facilitation Unit], Karnataka for supporting the research work.